\documentstyle[epsfig, twocolumn]{venice97}

\def\lessapprox{\,\raise 0.6ex\hbox{$<$}\kern -0.75em\lower 0.47ex
    \hbox{$\sim$}\,}

\def\href#1{#1}
\def\bibcode#1{}

\def\msun{{\rm M}_\odot}
\def\vbluw{$V\! BLUW$}

\def\aj#1{AJ\ #1}
\def\apj#1{ApJ\ #1}
\def\aap#1{A\&A\ #1}
\def\araa#1{ARA\&A\ #1}
\def\aapss#1{A\&AS\ #1}
\def\apjss#1{ApJS\ #1}

\def\mn#1{MNRAS\ #1}

\begin{document}

\setlength{\parindent}{0pt}
\setlength{\parskip}{ 10pt plus 1pt minus 1pt}
\setlength{\hoffset}{-1.5truecm}
\setlength{\textwidth}{ 17.1truecm }
\setlength{\columnsep}{1truecm }
\setlength{\columnseprule}{0pt}
\setlength{\headheight}{12pt}
\setlength{\headsep}{20pt}
\pagestyle{veniceheadings}

\title{\bf STRUCTURE AND EVOLUTION OF NEARBY OB ASSOCIATIONS}

\author{{\bf P.T.~de Zeeuw$^1$, 
             A.G.A.~Brown$^1$, 
             J.H.J.~de Bruijne$^1$, 
             R.~Hoogerwerf$^1$, 
             J.~Lub$^1$, 
             R.S.~Le Poole$^1$, 
             A.~Blaauw$^{1,2}$
             } \vspace{2mm} \\
$^1$Sterrewacht Leiden, P.O.~Box 9513, 2300 RA Leiden, The Netherlands \\
$^2$Kapteyn Astronomical Institute, P.O.~Box 800, 9700 AV Groningen, The Netherlands}

\maketitle

\begin{abstract}

We present the first results of a comprehensive census of the stellar
content of the nearby OB associations based on Hipparcos positions,
proper motions and parallaxes for 12842 candidate member stars
distributed over 21 fields on the sky. We use a new method to identify
moving groups in these fields (see de Bruijne et al., these
proceedings).  Previously, astrometric membership in nearly all the
nearby OB associations was known only for stars with spectral types
earlier than B5. The Hipparcos measurements now allow us to identify
members down to late F. This census provides a firm basis for studies
of galactic and extragalactic star forming regions.
\vspace {5pt} \\

Key~words: OB associations; HR diagram; luminosity calibration; stars:
early-type

\end{abstract}

\section{INTRODUCTION}

Ever since spectral classifications for the bright stars became
available it was evident that O and B stars are not distributed
randomly on the sky, but instead are concentrated in loose groups
(\cite{Blaauw64} and references therein). \cite*{Ambart47} found that
the stellar mass density in these groups, which were subsequently
called OB associations, is usually less than $0.1 \msun$~pc$^{-3}$.
\cite*{Bok34} had already shown that such low-density stellar groups
are unstable against Galactic tidal forces, so that the observed OB
associations must be young, a conclusion supported by the ages derived
from Hertzsprung--Russell diagrams. These groups are prime sites for
the study of star formation processes and of the interaction of
early-type stars with the interstellar medium (see, e.g., Blaauw 1964,
1991 for reviews).  Detailed knowledge of the stellar content and
structure of OB associations allows us to address fundamental
questions on the formation of stars in giant molecular clouds. What is
the initial mass function? What are the characteristics of the initial
binary population?  What is the star formation efficiency? Do all
stars in a group form at the same time?  What process causes the
distinction between the formation of bound open clusters and unbound
associations? How is angular momentum redistributed during star
formation?

The study of OB associations is also important in the context of the
evolution of the Galaxy. The kinematics of the nearest OB associations
provides insight into the properties and origin of the Gould Belt
system. Furthermore, OB associations are responsible for large bubbles
in the interstellar medium filled with hot gas (e.g.,
\cite{MacLow88}). More detailed knowledge of their stellar content is
important for understanding the energetics and dynamical evolution of
the bubbles.  Ultimately, establishing the properties of the nearby
associations is a prerequisite for the interpretation of observations
of extragalactic star forming regions and starburst galaxies.

\begin{figure*}[!ht]
%%%%%%%%%%%%%%%%%%%%
%
% Proper motions figure
%
%%%%%%%%%%%%%%%%%%%%
\begin{center}
\leavevmode
\centerline{\epsfig{file=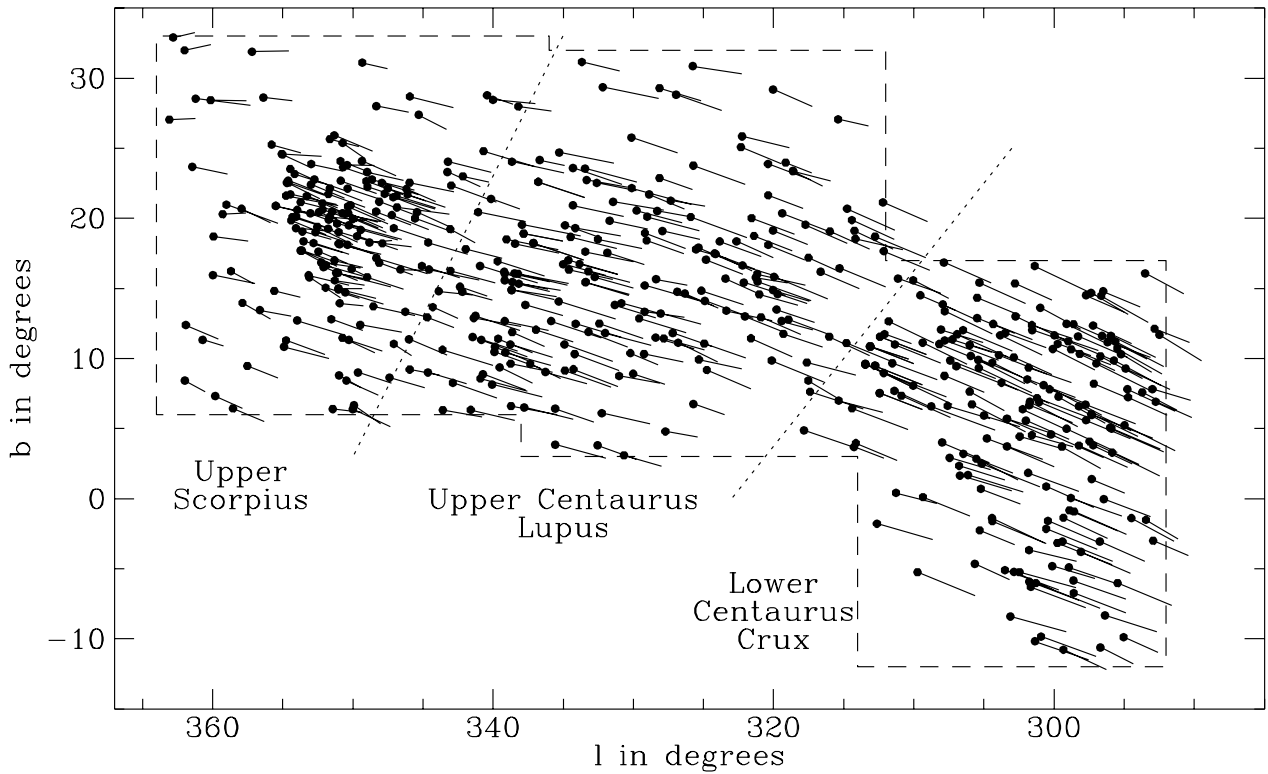,width=16.8cm}}
\end{center}
\caption{\em
Proper motions for 532 members of the Sco OB2 association selected
from 4156 candidate stars in our Hipparcos sample for the area bounded
by the dashed lines.  The dotted lines are the schematic boundaries
between the three classical subgroups Upper Scorpius, Upper Centaurus
Lupus, and Lower Centaurus Crux.
}
\label{fig:motions}
\end{figure*}

The Solar neighbourhood contains a number of OB associations, so that
detailed studies are possible. The most reliable membership
determinations for OB associations are based on proper motion studies.
Although these groups are unbound, their expansion velocities are only
a few~km s$^{-1}$ (e.g., \cite{Math86}), so that the common space
motion is perceived as a motion of the members towards a convergent
point on the sky (e.g., \cite{Blaauw46}; \cite{Bertiau58}). Membership
determinations have been carried out by various investigators (see
Blaauw 1964, 1991 and references therein). Due to their large extent
on the sky, OB associations are not amenable to proper motion studies
with photographic plates. Instead, one had to rely on proper motions
from large scale surveys with meridian circles. This resulted in much
uncertainty on membership of stars of spectral type later than B5.  As
a result, our knowledge of these young stellar groups remained rather
limited.

In order to remedy this problem the {\tt SPECTER} consortium was
formed in Leiden in 1982. It successfully proposed the observation by
Hipparcos of candidate members of nearby OB associations.  An
extensive program of ground-based observations was carried out in
anticipation of the release of the Hipparcos data (see
\cite{Zeeuw94}). This included Walraven (\vbluw) photometry
(\cite{Geus89}; \cite{Brown94}), mm and radio observations of the
interstellar medium surrounding the associations (\cite{Geus92};
\cite{Brown95}), and spectroscopy aimed at obtaining precise radial
and rotational velocities ({\cite{Verschue97}; \cite{BrownVer97}).
Theoretical work included the distinguishing features of the formation
of bound and unbound groups (\cite{Verschue89}), and the reliability
of the so-called kinematic ages of OB associations (\cite{Brown97a}).

Here we present the preliminary results of the {\tt SPECTER} census of
the nearby associations based on the Hipparcos proper motions and
trigonometric parallaxes. We have developed a new procedure to
identify moving groups in the Hipparcos Catalog (\cite{Bruijne97}),
and have applied it to 21 fields for which we have Hipparcos data. The
fields are centered on 18 known associations and 3 suspected groups,
all within 800 pc from the Sun. A list of the fields and their
boundaries, with the number of objects, can be found in
\cite*{Zeeuw94}. The dataset also includes 153 candidate runaway OB
stars as well as 49 stars from the dispersed Cas--Tau association.
Below we discuss a few examples of the results obtained so far. Other
examples are given in \cite*{Bruijne97} and \cite*{Hoogerwerf97}.

\begin{figure*}[!ht]
\begin{center}
\leavevmode
\centerline{\epsfig{file=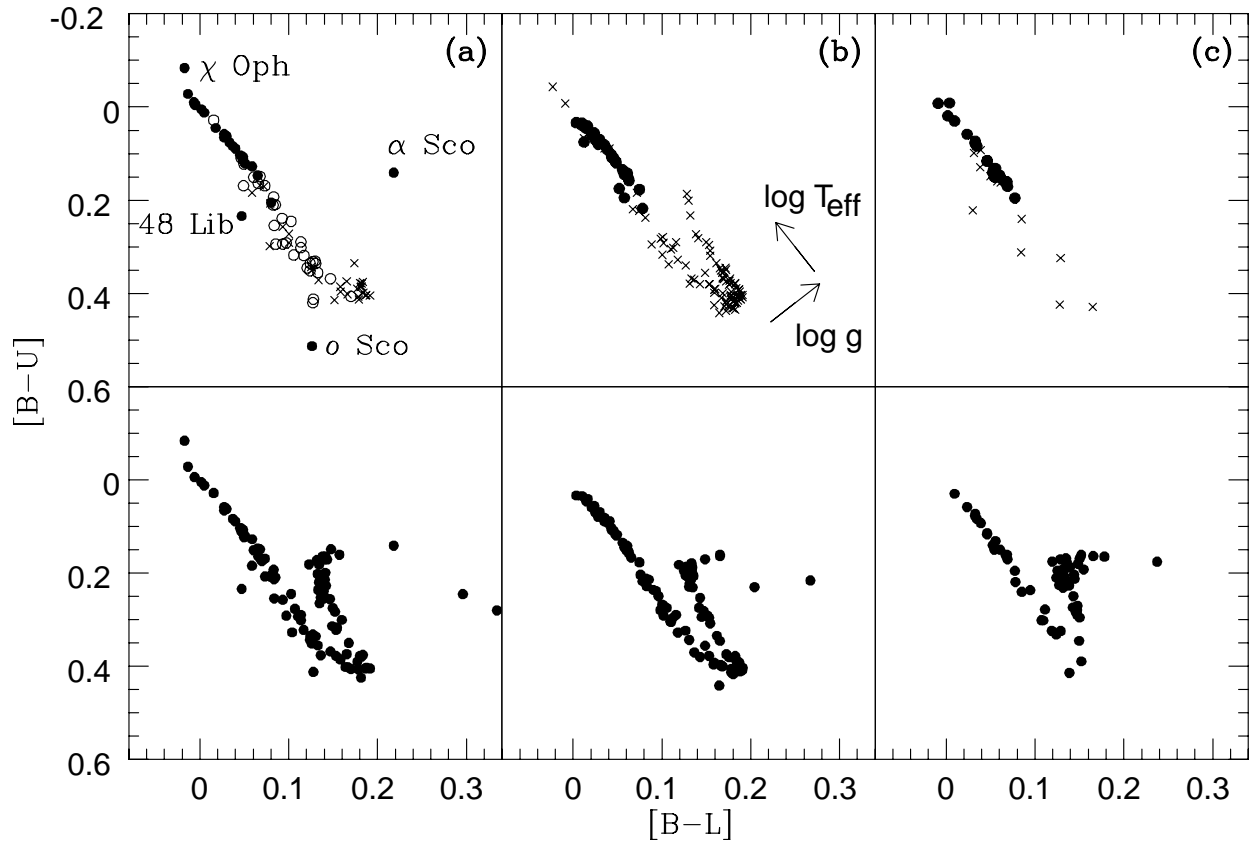,width=16.5cm}}
\end{center}
\caption{\em
De-reddened Walraven colour-colour diagrams for the three classical
subgroups of Sco OB2. a) 84/91 pre-Hipparcos `members' of Upper
Scorpius (top) and 106/178 new members (bottom) for which we have
Walraven photometry. Specific stars are indicated. b) Idem for Upper
Centaurus Lupus for 124/130 pre-Hipparcos `members' and 106/188 new
members, and c) for Lower Centaurus Crux for 34/42 pre-Hipparcos
`members' and 64/166 new members. The symbols in the top panels have
the following meaning: solid dots are the classical proper motion
members, the open circles are pre-Hipparcos probable proper motion
members and the crosses indicate previously suggested photometric
members. The arrows in (b) indicate the approximate directions of
increasing $\log T_{\rm eff}$ and $\log g$.
}
\label{fig:vbluw}
\end{figure*}

\section{SCORPIO--CENTAURUS}

The Sco OB2, or Scorpio--Centaurus, association contains three
classical subgroups which differ in age (\cite{Blaauw46},
\cite{Blaauw64}; \cite{Geus89}): Upper Scorpius, Upper Centaurus
Lupus, and Lower Centaurus Crux. Our membership selection for Upper
Scorpius is described in detail by \cite*{Bruijne97}. We have
similarly analysed the other two fields.  For Upper Centaurus Lupus 19
of the 29 classical proper motion members, and 30 of the 101
previously known photometric members were selected. We have discovered
another 139 members.  For Lower Centaurus Crux we confirm 15 of the
classical 22 proper motion members, and 3 of the 10 additional
photometric members. We have added another 148 members, including one
M giant. Figure~\ref{fig:motions} illustrates the motions in the three
fields, and shows that assigning boundaries to the subgroups is not
trivial. The numbers given above are based on the preliminary
boundaries used by \cite*{Zeeuw94}. Analysis of the entire Hipparcos
Catalog, and inclusion of radial velocity data, is needed to determine
the full extent of Sco OB2, and the division in subgroups.

Prior to the Hipparcos mission, we collected precise intermediate band
Walraven \vbluw\ photometry for 2243 of the 4156 Hipparcos stars in
the three Sco OB2 fields (\cite{Geus89}, 1990). We have measurements
for 276 of the total of 532 Hipparcos members. Figure~\ref{fig:vbluw}
shows the Walraven $[B - U]$ versus $[B - L]$ diagram for the
previously suspected members (top) and the Hipparcos members (bottom).
Figure~\ref{fig:vbluw} shows that the Hipparcos measurements allow
identification of new members to much later spectral types, all the
way to the mid F regime! The removal of interlopers in the
pre-Hipparcos membership lists results in a much tighter correlation
near the S-turn of the main sequence around $([B - L], [B - U])
\approx (0.2, 0.4)$. Stellar evolution moves stars towards the lower
left: this causes the slight downward curve at the top of the main
sequence. The remaining spread is caused by (i) peculiar spectra with,
e.g., emission features (cf.\ 48~Lib and $\chi$~Oph), (ii)
(undetected) duplicity/multiplicity; and (iii) stellar rotation: high
values of $v \sin i$ move stars towards the lower left. This effect is
strongest for B7 -- B9 stars ($0.05 \lessapprox [B - L] \lessapprox
0.13$).

The new membership lists can be used to derive reliable mass functions
to much smaller masses than possible previously. This requires
multi-colour photometry for all members, as well as a careful
correction for incompleteness in the Hipparcos Catalog at faint
magnitudes.  Application of our membership selection procedure to
simulated moving groups superimposed on the Galactic field population
shows that the expected number of interlopers is insignificant for
spectral types earlier than $\sim$ F5. De Geus et al.\ (1989) derived
ages of $\sim$ 5 Myr for Upper Scorpius, $\sim 10$ Myr for Lower
Centaurus Crux and $\sim 13$ Myr for Upper Centaurus Lupus. These
values are not expected to change much, as most of the previously
known bright members have been confirmed by our selection
procedure. The stars beyond spectral type F may not yet have had time
to reach the main sequence, and hence a detailed analysis of
membership for these late-type stars will be very interesting.

\vskip 1.0truecm

\begin{figure*}[!ht]
\begin{center}
\leavevmode
\centerline{\epsfig{file=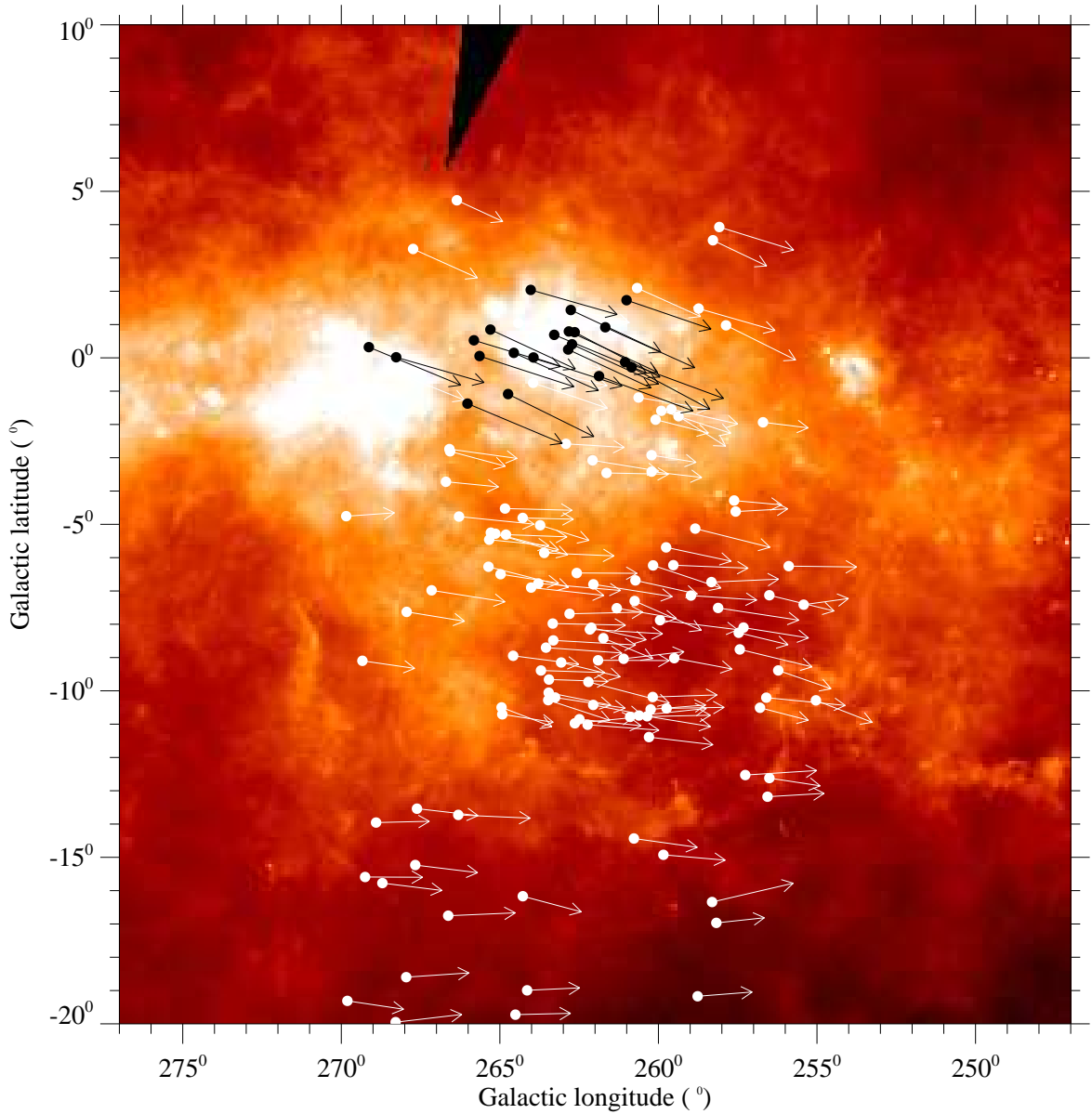,width=16.2cm,angle=90}}
\end{center}
\caption{\em
Proper motions in the Vela OB2 association superimposed on a
grey-scale representation of the IRAS 100~$\mu$m skyflux. The white
dots show the Vela OB2 members and the black dots show the members of
Trumpler 10 (located at a distance of $362\pm20$~pc). The core of the 
Vela OB2 association is surrounded by the IRAS Vela shell. 
}
\label{fig:vela}
\end{figure*}

\section{VELA OB2} 

\cite*{Brandt71} noted the presence of 17 bright early-type stars
within a few degrees of the Wolf--Rayet binary system $\gamma^2$
Velorum. They took the similar distance moduli for 10 of these stars
(including the $\gamma^2$ Vel system) as evidence for an OB
association at $\sim 460$~pc: Vela OB2. \cite*{Straka73} investigated
SAO proper motions and Bright Star Catalog radial velocities for these
10 stars.  Only 5 of them, including the visible multiple system
$\gamma^1$ and $\gamma^2$ Vel, turned out to share a common space
motion.  Photometric evidence for the existence of an association at a
distance of about 450 pc was provided by several large-scale studies
(e.g., \cite{Upton71}; \cite{Straka73}; \cite{Eggen86}).

Our Hipparcos sample for Vela OB2 contains 510 stars of spectral types
O (9) and B (501) in the field $255^\circ\! \leq \ell \leq
270^\circ\!$ and $-20^\circ\! \leq b \leq 5^\circ\!$.  The magnitude
limits of our sample are $V \leq 9.0$ for the O and B0--B5 stars, and
$V \leq 10.0$ for B6--B9 stars.  Only 4 of the 10 Brandt et al.\ stars
are confirmed as member.  Of the 6 remaining stars, $\gamma^1$ Vel was
not observed by Hipparcos, and the other 5 are rejected as
members. However, our membership selection procedure identifies 112
new members for Vela OB2, which brings the total number to 116: 1 O9V
star, $\gamma^2$ Vel, and 114 B-type stars!

We find a mean distance of $d = 415 \pm 10$ pc for Vela OB2. The
intrinsic depth of the association cannot be resolved from the
parallax distribution.  The new members of Vela OB2 are concentrated
on the sky around $(\ell, b) = (263^{\circ}, -7^{\circ})$ within a
radius of $\sim 5^{\circ}$. \cite*{Sahu92} reported the detection of
the so-called IRAS Vela shell in the IRAS Sky Survey Atlas maps. This
is an expanding shell, centered on Vela OB2, with a projected radius
of $\sim 8^{\circ}$ (Figure~\ref{fig:vela}). Sahu assumed that (i) the
center of the IRAS Vela shell has a distance of 460~pc, (ii) Vela OB2
is a `standard association' with a `normal' initial mass function, and
(iii) Vela OB2 has an age of 20 Myr, and she showed that then the
observed kinetic energy of the IRAS Vela shell is of the same order of
magnitude as the total amount of energy that the stars have injected
into the interstellar medium through the combined effects of stellar
winds and supernovae. Now that the present stellar content and the
distance of Vela OB2 have been determined, a careful multi-colour
photometric study will allow a considerable refinement of Sahu's
analysis. The Hipparcos photometry already indicates that many of the
Vela OB2 member stars are evolved.

The Vela OB2 field contains a number of other moving groups.  One
example, Trumpler 10, is illustrated in Figure~\ref{fig:vela}.

\section{OTHER GROUPS}

We have also detected moving groups in the fields of Per OB2
(\cite{Blaauw50}), Lac OB1 (\cite{Blaauw53}), and Cep OB3
(\cite{Garmany73}), and have confirmed the reality of the highly
dispersed association Cas--Tau (\cite{Blaauw56}). The associations
$\alpha$ Persei (Per OB3) and Collinder 121, as well as a new group
discovered in the field of Cep OB2, are discussed by
\cite*{Hoogerwerf97}. The Orion OB1 association has been studied
extensively, mostly by photometric means (e.g., \cite{Warren78};
\cite{Brown94}). It lies at a distance of $\sim 380$ pc, near the
direction of the Solar antapex. As a result, it is difficult to
distinguish its members from the general Galactic disk population
based only on the Hipparcos proper motions and parallaxes. We will
report on this interesting association elsewhere.

Our Hipparcos sample also includes 153 candidate OB runaway stars. The
study of these stars is of interest for settling the question of their
origin; supernova explosions in high mass binaries (\cite{Blaauw61})
or dynamical ejection (e.g., \cite{Gies86}). We are in the process of
retracing the paths of the runaways and the OB associations in the
Galactic potential in order to identify the parent associations, as
well as the age of the runaways. This is of considerable interest for
studies of high mass binary evolution (e.g., \cite{Rensbergen96}) and
of high mass X-ray binaries (e.g., \cite{Kaper97}).

\section{DISTANCES} 
 
We have used the Hipparcos parallaxes for the members in the
associations to determine their mean distances. This requires some
care, as the inverse of the parallax is a biased distance indicator
(\cite{Brown97b}).  As there is no evidence for a strong central
concentration of stars in our associations, we assume that to first
order the members are distributed homogeneously in a sphere. In this
case the expectation value of the mean of the measured parallaxes is
equal to the true parallax, and corresponds to the true distance of
the association.  The resulting distances to the associations are
presented in Table~\ref{tab:table}. The errors are derived from the
errors in the mean parallaxes.  Figure~\ref{fig:distances} shows that
the new distances to the associations are systematically smaller than
previous estimates which were based mostly on photometry.

Due to the selection effects in the Hipparcos Catalog, the observed
distribution of parallaxes may not be representative of the true
underlying parallax distribution of an association. For example,
magnitude limits will bias the sample towards the stars closest to us,
and therefore could be responsible for (part of) the discrepancy
evident in Figure~\ref{fig:distances}. For this reason we have carried
out Monte Carlo simulations of the distance determination of spherical
associations, which take into account the luminosity function, the
Hipparcos selection (in a crude approximation) and the error on the
parallax as a function of apparent magnitude. We find that the
magnitude limit bias is small, and conclude that the results presented
in Figure~\ref{fig:distances} are robust. Part of the difference in
distances is no doubt due to the greatly improved membership lists,
especially for the more distant associations. However, we suspect that
the calibration of the upper main sequence in the Hertzsprung--Russell
diagram may need revision.

\begin{table}[htb]
  \caption{\em Distances to OB associations. The first column lists
  the association and the second column the number of members
  identified from our Hipparcos sample. The third column contains the
  pre-Hipparcos estimate of the distance to these associations and the
  last column lists the distance derived from the mean Hipparcos
  parallax. The errors on the Hipparcos distances correspond to the
  errors in the mean parallax.}  
  \label{tab:table}
  \begin{center} \leavevmode 
  \footnotesize 
  \begin{tabular}[h]{lrcc}
  \hline \\[-5pt] Name & $N$ & $D_{\rm cl}$ (pc) & $D$ (pc)\\[+5pt]
  \hline \\[-5pt] Upper Scorpius & 178 & 160 & $145\pm\phantom{0}2$ \\
  Upper Centaurus Lupus & 188 & 145 & $140\pm\phantom{0}2$ \\ Lower
  Centaurus Crux & 166 & 120 & $118\pm\phantom{0}2$ \\ $\alpha$
  Persei/Perseus OB3 & 87 & 170 & $176\pm\phantom{0}5$ \\ Perseus OB2
  & 26 & 360 & $305\pm25$ \\ Trumpler 10 & 21 & 400 & $362\pm20$ \\
  Vela OB2 & 116 & 450 & $415\pm10$ \\ Lacerta OB1a & 45 & 530 &
  $382\pm25$ \\ Lacerta OB1b & 24 & 530 & $364\pm25$ \\ Collinder 121
  & 105 & 760 & $546\pm30$ \\ Cepheus OB2 & 49 & 790 & $559\pm30$ \\
  \hline\\ 
\end{tabular} 
\end{center}
\end{table}

\begin{figure}[!ht]
\begin{center}
\leavevmode
\centerline{\epsfig{file=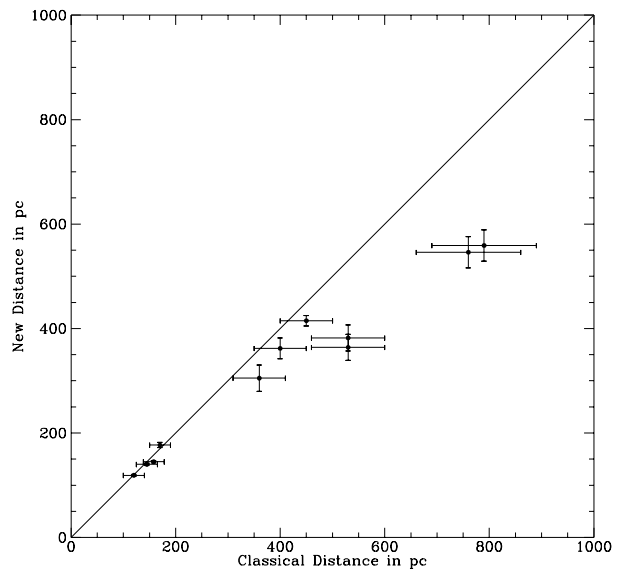,width=8cm}}
\end{center}
\caption{\em Distances derived from the mean parallax of the
Hipparcos members of OB associations versus previous best estimates of
the distance to these associations.}
\label{fig:distances}
\end{figure}

\section{CONCLUDING REMARKS} 

Application of our membership selection procedure to Hipparcos
parallaxes and proper motions for 12842 candidate stars in 21 fields
with previously known or suspected OB associations has significantly
improved our knowledge of their stellar content. The number of firmly
established kinematic members has increased dramatically.  The
membership lists for the early spectral types have been refined, and
many new members have been found of later spectral type, into and in
some cases beyond the regime of the F stars, which is the range where
the stars in these young groups are still in their pre-main sequence
phase.

The measured distances for the associations are systematically smaller
than indicated by previous photometric determinations. Whereas part of
this effect must be caused by the much improved membership lists,
especially for the more distant groups, we suspect that a
recalibration is needed of the upper main sequence in the
Hertzsprung--Russell diagram.

The next step is to re-examine our 21 fields, and the adjacent
regions, using all data contained in the Hipparcos Catalog. We expect
to find additional members that were not included because of our
magnitude limits or field boundaries. The resulting astrometric
membership lists can then be refined further by radial velocity
measurements. This is now feasible, as the astrometric membership
selection has reduced the number of candidate association members in a
field typically by an order of magnitude. This also applies to the
required completion of the multi-colour photometry. In the near future
we will obtain radial velocities for the associations Lac OB1, Cep OB2
and Per OB2 from approved observations at the McDonald and
Haute-Provence observatories. We will study the binary population in
Sco OB2 by means of radial velocity measurements and near-infrared
adaptive optics searches for close companions to the B-type stars.
Subsequently we will extend these studies to the other associations.

\section*{ACKNOWLEDGMENTS}

It is a pleasure to thank Eug\`{e}ne de Geus and Michael Perryman
for stimulating discussions, and Jan Brand and
Frank Israel for significant contributions to the 1982 {\tt SPECTER}
proposal.

\end{document}